\documentclass[a4paper]{amsart}
\usepackage{amsmath, amssymb, amsfonts, amscd}

\numberwithin{equation}{section}
\theoremstyle{plain}
 \newtheorem{thm}{Theorem}[section]
 \newtheorem{cor}[thm]{Corollary}
 \newtheorem{lem}[thm]{Lemma}
 \newtheorem{prop}[thm]{Proposition}
\theoremstyle{definition}
 \newtheorem{defn}[thm]{Definition}

\theoremstyle{remark}
 \newtheorem{rem}[thm]{Remark}

\DeclareMathOperator{\sh}{sinh}
\DeclareMathOperator{\ch}{cosh}
\DeclareMathOperator{\ord}{ord}

\def\p{\partial}
\begin{document}
\title[completely integrable quantum systems]%
{A class of completely integrable quantum systems associated with 
classical root systems}%
\author{Toshio Oshima}
\address{Graduate School of Mathematical Sciences,
University of Tokyo, 7-3-1, Komaba, Meguro-ku, Tokyo 153-8914, Japan}
\email{oshima@ms.u-tokyo.ac.jp}
\dedicatory{Dedicated to Professor Gerrit van Dijk on the occasion of 
his sixty fifth birthday}
\begin{abstract}
We classify the completely integrable systems associated with
classical root systems whose potential functions are meromorphic at an 
infinite point.
\end{abstract}
\maketitle
\section{Introduction}
A Shr\"odinger operator 
\begin{equation}
  P=\sum_{j=1}^n\frac{\p^2}{\p x_j^2} + R(x)\label{eq:Shr}
\end{equation}
with the potential function $R(x)$ of $n$ variables $x=(x_1,\dots,x_n)$ 
is called {\it completely integrable\/} if
there exist $n$ differential operators $P_1,\dots,P_n$ such that
\begin{equation}\label{eq:CI}
 \begin{cases}
   [P_i,P_j]=0\qquad(1\le i < j\le n),\\
   P\in\mathbb C[P_1,\dots,P_n],\\
   P_1,\dots,P_n\text{ are algebraically independent}.
 \end{cases}
\end{equation}
In this note $P$ is called to be completely integrable {\it of type $B_n$\/} 
or {\it of classical type\/}
if $P_k$ and $R(x)$ in the above are of the forms
\begin{align}
 P_k &= \sum_{j=1}^n\frac{\p ^{2k}}{\p x_j^{2k}}+ Q_k\qquad
 \text{with }\ord Q_k<\ord P_k,\label{eq:Bn}\\
 R(x) &= \sum_{1\le i<j\le n}\Bigl(u^-_{ij}(x_i-x_j)+u^+_{ij}(x_i+x_j)\Bigr)
        +\sum_{k=1}^n v_k(x_k).\label{eq:1var}
\end{align}
Here $u^\pm_{ij}$ and $v_k$ are functions of one variable.

The systems of differential operators satisfied by 
the radial parts of zonal spherical functions or Whittaker functions on 
Riemannian symmetric spaces of the non-compact and classical type, 
Heckman-Opdam's hypergeometric equations (cf.~\cite{HO}), 
 Calogero-Moser and Sutherland systems for
one dimensional quantum $n$-body problems (cf.~\cite{OP1}, \cite{OP2})
and Toda finite chains associated with (extended) classical Dynkin diagrams
are their examples.

We remark that \cite{Wa} proves that if the potential function 
$R(x)$ is locally defined and analytic, then the condition \eqref{eq:CI} 
with \eqref{eq:Bn} assures \eqref{eq:1var}
and moreover $R(x)$ is extended to a global meromorphic function on 
$\mathbb C^n$ except for a trivial case corresponding to Type $A_1$ in 
Theorem~\ref{prop:bn}
(cf.\ \cite{Oc} for type $B_2$ and \cite{OS} in the invariant case).

\cite{OOS}, \cite{OS}, \cite{O2} and \cite{OO} determine this integrable 
system under the condition that $P$ is $B_n$-invariant, 
namely, $u^+_{ij}$, $u^-_{ij}$ and $v_k$ are even functions and do not
depend on $i$, $j$ and $k$ and $u^+_{ij}=u^-_{ij}$. 
On the other hand
\cite{Oc}, \cite{Ta} and \cite{Wa} determine it if $R(x)$ has certain
singularities.

We assume in this note
that $R(x)$ is meromorphic at $t=0$ under the coordinate system
\begin{equation}\label{eq:coord}
 t_j = e^{-(x_j-x_{j+1})}\quad(j=1,\ldots,n-1),\quad t_n = e^{-x_n}
\end{equation}
and classify the Shr\"odinger operator \eqref{eq:Shr} 
which allows a differential operator $P_2$ of the form \eqref{eq:Bn} satisfying
$PP_2= P_2P$.
We note that the above examples with non-rational potential functions
satisfy this assumption.
In the first example this follows from the fact that the 
invariant differential operators on a Riemannian symmetric space have 
analytic extensions on a smooth compactification of the space
(cf.~\cite{O1}).

Theorem~\ref{prop:bn} and Remark~\ref{rm:bn} in \S\ref{sec:Bn} determine 
$R(x)$, which is the main result of this note and proved by using
\S\ref{sec:An} and \S\ref{sec:B2}.
The result implies that the system is a suitable
limit of the invariant quantum integrable system classified by \cite{OOS}
(cf.~\cite{I}, \cite{IM}, \cite{Ru}, \cite{vD} and \cite{vD2}).
Hence the integrals $P_1,\dots,P_n$ will be calculated as suitable
limits from the integrals 
given in \cite{O2}, which 
will be shown in another paper.

If $R(x)$ is analytic at $t=0$, we say that $R(x)$ has 
{\it regular singularity\/} at the infinite point $t=0$, which are also
classified in Corollary~\ref{cor:regB}.

In \S\ref{sec:B2} the potential function $R(x)$ is determined when $n=2$.

In \S\ref{sec:An} we study the potential function $R(x)$ when 
$u^+_{ij}=v_k=0$, which we call to be of type $A_{n-1}$.

\section{Type $A_{n-1}$ ($n\ge3$)}\label{sec:An}
In this section we study the Shr\"odinger operator
\begin{equation}
 P=\sum_{j=1}^n\frac{\p^2}{\p x_j^2} + \sum_{1\le i<j\le n} 
 \tilde u_{ij}(x_i-x_j)
\end{equation}
which allows a differential operator
\begin{equation}
 Q=\sum_{1\le i<i<k\le n}\frac{\p^3}{\p x_i\p x_j\p x_k}+S
 \qquad\text{with }\ord S<3
\end{equation}
satisfying $[P,Q]=[\sum_{j=1}^n\frac{\p}{\p x_j},Q]=0$.
Then the proof of \cite[Proposition 4.2]{OOS} implies that the
existence of $Q$ is equivalent to 
\begin{equation}\label{eq:an}
  \sum_{1\le i<j<k\le n}U_{ijk} = 0
\end{equation}
with
\begin{align}
\begin{aligned}
  U_{ijk} &= 
     u_{jk}(t_j\cdots t_{k-1})\Bigl(t_i\cdots t_{j-1}u'_{ij}(t_i\cdots t_{j-1})
     + t_i\cdots t_{k-1}u'_{ik}(t_i\cdots t_{k-1})\Bigr)\\
     &+ u_{ik}(t_i\cdots t_{k-1})\Bigl(-t_i\cdots t_{j-1}u'_{ij}(t_i\cdots t_{j-1})
     + t_j\cdots t_{k-1}u'_{jk}(t_j\cdots t_{k-1})\Bigr)\\
     &-u_{ij}(t_i\cdots t_{j-1})\Bigl(t_i\cdots t_{k-1}u'_{ik}(t_i\cdots t_{k-1})
     + t_j\cdots t_{k-1}u'_{jk}(t_j\cdots t_{k-1})\Bigr)
\end{aligned}
\end{align}
by putting $u_{ij}(e^{-y}) = \tilde u_{ij}(y)$.
We assume that $R(x)$ is holomorphic for $0<|t|\ll1$ under the coordinate system \eqref{eq:coord} which corresponds to the expression
\begin{equation}
   u_{ij}(s) = \sum_{\nu\in\mathbb Z}c^{ij}_\nu s^\nu\quad(c^{ij}_0=0)
  \text{ converge for }0<|s|\ll 1.
\end{equation}
We assume $c^{ij}_0=0$ without loss of generality and
expand \eqref{eq:an} into the power series. 
Then the terms $(t_i\cdots t_{j-1})^p(t_j\cdots t_{k-1})^q$ with $p\ne0$, 
$q\ne 0$, $p\ne q$ and $i<j<k$ appear only in $U_{ijk}$ and therefore if 
$p\ne0$, $q\ne 0$ and $p\ne q$, we have
\[
  c^{jk}_q pc^{ij}_p + c^{jk}_{q-p}pc^{ik}_p
  -c^{ik}_q(p-q)c^{ij}_{p-q}+c^{ik}_p(q-p)c^{jk}_{q-p}
  -c^{ij}_{p-q}qc^{ik}_q - c^{ij}_pqc^{jk}_q=0
\]
and hence
\[
 (p-q)c^{ij}_pc^{jk}_q - pc^{ij}_{p-q}c^{ik}_q + qc^{jk}_{q-p}c^{ik}_p = 0.
\]
Denoting
\begin{equation}\label{eq:anU}
 U_{ij}(t) = \sum_{\nu\in\mathbb Z\setminus\{0\}}C^{ij}_\nu t^\nu\quad
 \text{with } c^{ij}_\nu = \nu C^{ij}_\nu,
\end{equation}
we have
\begin{gather}
 u_{ij}(t) = tU'_{ij}(t),\\
 pq(p-q)\bigl(C^{ij}_pC^{jk}_q - C^{ij}_{p-q}C^{ik}_q - C^{jk}_{q-p}C^{ik}_p\bigr) 
 = 0.\label{eq:ani}
\end{gather}
Then \eqref{eq:an} is equivalent to
\begin{equation}\label{eq:a2f}
 \bigl(U_{ij}(s) + U_{jk}(t) - U_{ik}(st)\bigr)^2 = 
 V_{ij}^{ijk}(s) + V_{jk}^{ijk}(t) - V_{ik}^{ijk}(st)
\end{equation}
with suitable functions $V_{ij}^{ijk}$, $V_{jk}^{ijk}$ and $V_{ik}^{ijk}$
for $1\le i<j<k\le n$.

\begin{rem}\label{rem:a2}
If $\bigl(U_{ij}(t), U_{jk}(t), U_{ik}(t)\bigr)$ satisfies \eqref{eq:a2f} 
with suitable $V_{ij}$, $V_{jk}$ and $V_{ij}$, then
$\bigl(U_{jk}(t), U_{ij}(t), U_{ik}(t)\bigr)$ and
$\bigl(cU_{ij}(at^r), cU_{jk}(bt^r), cU_{ik}(abt^r)\bigr)$ have the same 
property for any complex numbers $a$, $b$ and $c$ and a positive integer $r$ 
with $ab\ne 0$.
\end{rem}

\begin{prop}\label{prop:A}
The solution $(U_{ij}, U_{jk}, U_{ik})$ of \eqref{eq:a2f} with \eqref{eq:anU}
is one of the followings and it satisfies
$U_{ijk}=0$.

{\rm i)}
Two of $\{U_{ij}, U_{jk}, U_{ik}\}$ are zero and the other one is any function.

{\rm ii)}
$(U_{ij}, U_{jk}, U_{ik}) = (at^r, bt^r, ct^{-r})$ for any
$a$, $b$ and $c\in\mathbb C$ and $r\in\mathbb Z\setminus\{0\}$.

{\rm iii)}
$(U_{ij}, U_{jk}, U_{ik}) = 
\bigl(\dfrac{act^r}{1-at^r}, \dfrac{bct^r}{1-bt^r}, \dfrac{abct^r}{1-abt^r}\bigr)$
with any non-zero complex numbers $a$, $b$ and $c$ and a positive integer $r$.
\end{prop}
\begin{proof}
All the solutions of the equation \eqref{eq:a2f} are obtained by 
\cite{BP} and \cite{BB} (cf.~\cite[Remark~2.3]{OO}), which implies this proposition.
But we will give a simple proof under the assumption that
the origin is at most a pole of $U_{ij}$, $U_{jk}$ and $U_{ik}$.

Suppose one of $U_{ij}, U_{jk}, U_{ik}$ is zero 
and the other two are not zero. 
If $U_{ij}=0$ and $C^{jk}_r\ne 0$, 
then we have $p(m+p)mC^{jk}_mC^{ik}_p=0$ and 
therefore $C^{ik}_p=0$ for $p\ne -r$, $C^{ik}_{-r}\ne 0$ and $C^{jk}_m=0$ for $m\ne r$.
Thus we have {\rm ii)}.  
We similarly have {\rm ii)} in the other two cases.

Hence we may assume that any one of $\{U_{ij}, U_{jk}, U_{ik}\}$ is not zero.
Define $I_{\ell m}\in\mathbb Z\setminus\{0\}$ such that $C^{\ell m}_{I_{\ell m}}\ne0$
and $C^{\ell m}_\nu = 0$ for $\nu< I_{\ell m}$.  Then \eqref{eq:ani} shows
\begin{equation}\label{eq:a2C}
 I_{ij}I_{jk}(I_{ij}-I_{jk})(C^{ij}_{I_{ij}}C^{jk}_{I_{jk}}
 - C^{ij}_{I_{ij}-I_{jk}}C^{ik}_{I_{jk}} - C^{jk}_{I_{jk}-I_{ij}}C^{ik}_{I_{ik}}) = 0.
\end{equation}

Suppose $I_{ij}>0$ and $I_{jk}>0$.  Then \eqref{eq:a2C} means $I_{ij}=I_{jk}$, which
we put $r$, and therefore \eqref{eq:ani} with $q=r$ and that with $p=q+r$ mean
\begin{align}
  pr(p-r)(C^{ij}_pC^{jk}_r - C^{ij}_{p-r}C^{ik}_r) &= 0
    \quad\text{for }p>0,\label{eq:a2sh}\\
  (q+r)qr(C^{ij}_{q+r}C^{jk}_q - C^{ij}_rC^{ik}_q) &= 0,\label{eq:a2she}
\end{align}
respectively.

If $C^{ik}_r=0$, it follows from \eqref{eq:a2sh} that $C^{ij}_p = 0$ for $p\ne r$
by the induction on $p$ and we have similarly $C^{jk}_q = 0$ for $q\ne r$ 
by the symmetry between $U^{ij}$ and $U^{jk}$ 
and finally $C^{ik}_q = 0$ for $q\ne -r$ by \eqref{eq:a2she}.
Hence this case is reduced to ii) with $r>0$.

Suppose $C^{ik}_r\ne 0$.  Then by Remark~\ref{rem:a2} we may assume 
$C^{ij}_r=C^{jk}_r=C^{ik}_r$ by a suitable transformation $(s,t)\mapsto(at,bt)$
and moreover by \eqref{eq:a2sh} that $U_{ij}=c\sum_{\nu=1}^\infty t^{r\nu}$
and similarly $U_{jk}=c'\sum_{\nu=1}^\infty t^{r\nu}$.
Then \eqref{eq:a2she} means $U_{ik} = U_{jk} + c''t^{-r}$.
Finally we have $c''=0$ by \eqref{eq:a2C} with $p=2r$ and $q=-r$
and get $U_{ij}=U_{jk}=U_{ik}$. 

Lastly we may assume $I_{ij}<0$ by Remark~\ref{rem:a2}.
Then \eqref{eq:ani} with $p=I_{ij}+I_{ik}$ and $q=I_{ik}$ implies $I_{ik}>0$ and
that with $p=I_{ij}$ and $q>0$ means $C^{jk}_q=0$ for $q\ge 0$.
Hence $I_{jk}<0$ and similarly we have $C^{ij}_p=0$ for $p\ge 0$.
Moreover \eqref{eq:ani} with $p=q+I_{ij}$ shows $C^{ik}_q=0$ for sufficiently 
large integer $q$.
Then $\bigl(U_{ij}(t^{-1}),U_{jk}(t^{-1}), U_{ik}(t^{-1})\bigr)$ 
is also a solution 
of \eqref{eq:a2f} and this case is reduced to the case when $I_{ij}>0$ and 
$I_{jk}>0$ and therefore we have ii) with $r<0$.

Note that it is easy to see that the given functions in the proposition 
satisfy $U_{ijk}=0$ (cf.~Remark~\ref{rem:a2}).
\end{proof}
\begin{rem}  If $t=e^{-x}$, then
\begin{align*}
 t\frac d{dt}\bigl(at^r\bigr) &= art^r=are^{-rx},\\
 t\frac d{dt}\Bigl(\frac{at^r}{1-at^r}\Bigr) &=
 \frac{art^r}{(1-at^r)^2} = r\sh^{-2}\frac{rx-\log a}2.
\end{align*}
\end{rem}

\section{Type $B_2$}\label{sec:B2}
In this section we study the following commuting differential operators.
\begin{equation}\label{eq:b2-0}
\begin{cases}
 P=\dfrac{\p ^2}{\p x^2}+\dfrac{\p ^2}{\p y^2} + R(x,y),\\
 Q=\dfrac{\p^4}{\p x^2\p y^2} + S \quad\text{ with }\ord S<4,\\
 [P,Q]=0.
\end{cases}
\end{equation}
Note that $P_2=P^2-2Q$ in \eqref{eq:Bn}.
First we review the arguments given in \cite{OO} and \cite{Oc}.
Since $P$ is self-adjoint, we may assume $Q$ is also self-adjoint by replacing
$Q$ by its self-adjoint part if necessary.
Here for $A=\sum a_{ij}(x,y)\frac{\p^{i+j}}{\p x^i\p y^j}$ we define
${}^t\!A=\sum (-1)^{i+j}\frac{\p^{i+j}}{\p x^i\p y^j}a_{ij}(x,y)$
and $A$ is called self-adjoint if ${}^t\!A=A$.
Then
\begin{equation}\label{eq:b2-1}
\begin{split}
  R(x,y) &= u^+(x+y)+u^-(x-y)+v(x)+w(y),\\
  Q&=\left(\frac{\p ^2}{\p x\p y} + \frac{u^+(x+y) - u^-(x-y)}2\right)^2 + 
  w(y)\frac{\p^2}{\p x^2} + v(x)\frac{\p ^2}{\p y^2}\\
   &\quad
   + v(x)w(y) + T(x,y),\\
\end{split}
\end{equation}
and the function $T(x,y)$ satisfies
\begin{equation}\label{eq:b2-2}
\begin{split}
  2\frac{\p T(x,y)}{\p x} &= \bigl(u^+(x+y)-u^-(x-y)\bigr)\frac{\p w(y)}{\p y}
    + 2w(y)\frac{\p}{\p y}\bigl(u^+(x+y)-u^-(x-y)\bigr),\\
  2\frac{\p T(x,y)}{\p y} &= \bigl(u^+(x+y)-u^-(x-y)\bigr)\frac{\p v(x)}{\p x}
    + 2v(x)\frac\p{\p x}\bigl(u^+(x+y)-u^-(x-y)\bigr).
\end{split}
\end{equation}
On the other hand, if a function $T(x,y)$ satisfies \eqref{eq:b2-2}
for suitable functions $u^\pm(t)$, $v(t)$ and $w(t)$, then \eqref{eq:b2-0} is
valid for $R(x,y)$ and $Q$ defined by \eqref{eq:b2-1}.

We have the compatibility condition 
\begin{equation}\label{eq:b2comp}
\begin{aligned}
\frac{\p}{\p x}\Bigl(&\bigl(u^+(x+y)-u^-(x-y)\bigr)\frac{\p v(x)}{\p x}
  + 2v(x)\frac\p{\p x}\bigl(u^+(x+y)-u^-(x-y)\bigr)\Bigr)\\
=&
\frac\p{\p y}\Bigl(\bigl(u^+(x+y)-u^-(x-y)\bigr)\frac{\p w(y)}{\p y}
  + 2w(y)\frac\p{\p y}\bigl(u^+(x+y)-u^-(x-y)\bigr)\Bigr)
\end{aligned}
\end{equation}
for the existence of $T(x,y)$.

\begin{defn}[Duality in $B_2$]\label{defn:dual}
Under the coordinate transformation
\begin{equation}
 (x,y)\mapsto\left(\frac{x+y}{\sqrt 2},\frac{x-y}{\sqrt 2}\right)
\end{equation}
the pair $(P,\frac14 P^2 - Q)$ also satisfies \eqref{eq:b2-0},
which we call the {\it duality\/} of the commuting differential operators of 
type $B_2$.
\end{defn}

Denoting $\p_x=\frac\p{\p x}$, $\p_y=\frac\p{\p y}$ and put 
\[
 L=P^2 - 4Q - (\p_x^2-\p_y^2 +v(x)-w(y))^2 - 2u^-(x-y)(\p_x+\p_y)^2 - 2u^+(x+y)(\p_x-\p_y)^2.
\]
Then the order of $L$ is at most 2 and the second order term of $L$ equals
\begin{multline*}
  2(u^++u^-+v+w)(\p_x^2+\p_y^2) - 4(u^+-u^-)\p_x\p_y - 4w\p_x^2 - 4v\p_y^2\\
 - 2(v-w)(\p_x^2-\p_y^2) -2u^-(\p_x+\p_y)^2-2u^+(\p_x-\p_y)^2 = 0.
\end{multline*}
Since $L$ is self-adjoint, $L$ is of order at most 0 and the 0-th order term of $L$
equals
\begin{multline*}
 (\p_x^2+\p_y^2)(u^++u^-+v+w) + (u^++u^-+v+w)^2 - 4(vw+T) - 2\p_x\p_y(u^+-u^-)\\
 -(\p_x^2-\p_y^2)(v-w) = (u^++u^-+v+w)^2 - 4(vw+T)
\end{multline*}
and therefore we have the following proposition.
\begin{prop} {\rm i)}
By the duality in Definition~\ref{defn:dual} the pair $\bigl(R(x,y),T(x,y)\bigr)$ changes into
$\bigl(\tilde R(x,y),\tilde T(x,y)\bigr)$ with
\begin{align}
 \begin{cases}
   \tilde R(x,y) = v\Bigr(\dfrac{x+y}{\sqrt2}\Bigl) + w\Bigl(\dfrac{x-y}{\sqrt2}\Bigr) 
                + u^+\bigl(\sqrt{2}x\bigr) + u^-\bigl(\sqrt{2}y\bigr),\\
   \tilde T(x,y) 
             = \dfrac14\tilde R(x,y)^2
             - v\Bigl(\dfrac{x+y}{\sqrt2}\Bigr)w\Bigl(\dfrac{x-y}{\sqrt2}\Bigr)
             - T\Bigl(\dfrac{x+y}{\sqrt2},\dfrac{x-y}{\sqrt2}\Bigr).
 \end{cases}
\end{align}

{\rm ii)} Combining the duality with the scaling map $R(x,y)\mapsto c^{-2}R(cx,cy)$, 
the following pair $\bigl(R^d(x,y),T^d(x,y)\bigr)$ defines commuting differential operators 
if so is $\bigl(R(x,y),T(x,y)\bigr)$.  
This $R^d(x,y)$ is also called the {\it dual\/} of $R(x,y)$.

\begin{align}
 \begin{cases}
   R^d(x,y) =  v(x+y) + w(x-y) + u^+(2x) + u^-(2y),\\
   T^d(x,y) = \frac14 R^d(x,y)^2 - v(x+y)w(x-y) - T(x+y,x-y).
 \end{cases}
\end{align}
\end{prop}
Now we give a list of the solutions of \eqref{eq:b2comp} and \eqref{eq:b2-2}.
They are suitable limits of the invariant solutions studied in \cite{OO}
and many of them are given in \cite{Oc}.

Case I: (Any-$A_1$)$+$(Any-$A_1$)\quad 
    $v=w=0$ and $u$ and $v$ are arbitrary functions.

Case II: $u^+=u^-$, $v=w$ and $(u^+;v)$ is in the following list.
\begin{gather*}
  (\langle \sh^{-2}\lambda t\rangle;\ \langle \sh^{-2}2\lambda t,\ 
   \sh^{-2}\lambda t,\ \ch2\lambda t,\ \ch4\lambda t\rangle),\label{eq:list-top}
   \tag{\rm Trig-$B_2$}\\ 
  (\langle \sh^{-2}\lambda t,\ \sh^{-2}2\lambda\rangle;\ 
   \langle\sh^{-2}2\lambda t,\ \ch 4\lambda t\rangle)
  \tag{\rm Trig-$B_2$-S}.
\end{gather*} 

Case III: $u^+=u^-$, $(u^+;v,w)$ is in the following list.
\begin{gather*}
  (\langle \ch2\lambda t\rangle;\ 
   \langle \sh^{-2}\lambda t,\ \sh^{-2}{2\lambda t}\rangle,\  
   \langle \sh^{-2}\lambda t,\ \sh^{-2}{2\lambda t}\rangle)
  \tag{\rm Toda-$D_2^{(1)}$-bry},\\
  (\langle \ch\lambda t,\ \ch2\lambda t\rangle;\ 
   \langle \sh^{-2}\lambda t\rangle,\  \langle \sh^{-2}\lambda t\rangle)
  \tag{\rm Toda-$D_2^{(1)}$-S-bry},\\
  (\langle e^{-2\lambda t}\rangle;\ 
   \langle e^{2\lambda t},\ e^{4\lambda t}\rangle,\ 
   \langle \sh^{-2}\lambda t,\ \sh^{-2}{2\lambda t}\rangle)
  \tag{\rm Toda-$B_2^{(1)}$-bry},\\
  (\langle e^{-\lambda t},\ e^{-2\lambda t}\rangle;\  
   \langle e^{2\lambda t}\rangle,\ \langle \sh^{-2}\lambda t\rangle)
  \tag{\rm Toda-$B_2^{(1)}$-S-bry}.
\end{gather*}

Case IV: $v=w$, $(u^+,u^-;v)$ is in the following list.
\begin{gather*}
  (0,\ \langle\sh^{-2}\lambda t\rangle;\ 
  \langle e^{-2\lambda t}, e^{-4\lambda t}, e^{2\lambda t}, e^{4\lambda t}\rangle)
  \tag{\rm Trig-$A_1$-bry},\\
  (0,\ \langle \sh^{-2}\lambda t,\ \sh^{-2}{2\lambda t}\rangle;\ 
   \langle e^{-4\lambda t},\ e^{4\lambda t}\rangle)
  \tag{\rm Trig-$A_1$-S-bry}.
\end{gather*}

Case V: $(u^+,u^-,v,w)$ is in the following list.
\begin{gather*}
  (0,\ \langle e^{-\lambda t}\rangle,\ 
   \langle e^{\lambda t},\ e^{2\lambda t}\rangle,\ 
   \langle e^{-\lambda t},\ e^{-2\lambda t}\rangle)
  \tag{\rm Toda-$C_2^{(1)}$},\\
  (0,\ \langle e^{-\lambda t},\ e^{-2\lambda t}\rangle,\ 
   \langle e^{2\lambda t}\rangle,\ 
   \langle e^{-2\lambda t}\rangle)
   \tag{\rm Toda-$C_2^{(1)}$-S}.
   \label{eq:list-last}
\end{gather*}
In the above $\langle\ \rangle$ means an arbitrary linear combination of 
given functions and, for example, (Trig-$B_2$) implies
\[
 \begin{cases}
 u^+(t)=u^-(t)=C_0\sh^{-2}\lambda t,\\ 
 v(t)=w(t)=C_1\sh^{-2}2\lambda t+C_2\sh^{-2}\lambda t
  +C_3\ch2\lambda t+C_4\ch4\lambda t
 \end{cases}
\]
with any complex numbers $C_0,C_1,\dots,C_4$ and 
a suitable $\lambda\in\mathbb C\setminus\{0\}$.

According to our assumption, put
\begin{equation}\label{eq:series}
\begin{gathered}
  t = e^{-y},\quad s = e^{-x+y},\\
  u^+(x+y) = \sum_{i\ge r}u^+_i s^it^{2i},\ 
  u^-(x-y) = \sum_{i\ge r}u^-_is^i,\\
  v(x) = \sum_{j\ge r'}v_j s^jt^j,\ 
  w(y) = \sum_{j\ge r''}w_jt^j,\\
  u^{\pm}_i=v_j=w_k=0\quad\text{if } i<r,\, j<r' \text{ and }k<r''.
\end{gathered}
\end{equation}
\begin{equation}\label{eq:exp1}
\begin{aligned}
  {}&\sum_{\substack{ i\ge r\\j\ge r'}}
       (i+j)(2i+j)v_j(u_i^+t^{2i+j} - u_i^-t^{j})s^{i+j}\\
  {}&= \sum_{\substack{i\ge r\\j\ge r''}}
       \Big((2i+j)(i+j)w_ju_i^+t^{2i+j}-(2i-j)(i-j)w_ju_i^-t^{j}\Big)s^i
\end{aligned}
\end{equation}
and the coefficients of $s^pt^q$ mean
\begin{equation}\label{eq:b2i0}
   pqv_{2p-q}u^+_{q-p} - p(2p-q)v_qu^-_{p-q}
 = q(q-p)w_{q-2p}u^+_p - (2p-q)(p-q)w_qu^-_p.
\end{equation}
Putting
\begin{equation}\label{b2:series2}
  \begin{cases}
    U^\pm(t) = \sum_{i\ge r}U^\pm_it^i,\ 
    V(t) = \sum_{j\ge r'}V_jt^j \text{ and }
    W(t) = \sum_{k\ge r''}W_kt^k,\\
    u^\pm(t)=t(U^\pm)'(t)+u^\pm_0,\ v(t)=tV'(t)+v_0
    \text{ and }w(t)=tW'(t)+w_0,
  \end{cases}
\end{equation}
we have
\begin{equation}\label{eq:b2i}
  pq(2p-q)(p-q)\bigl(V_{2p-q}U^+_{q-p} + V_qU^-_{p-q}
   + W_{q-2p}U^+_p - W_qU^-_p\bigr) = 0,
\end{equation}
which is equivalent to
\begin{multline}\label{eq:b2fun}
 V(st)\bigl(U^+(st^2) + U^-(s)\bigr) + 
 W(t)\bigl(U^+(st^2) - U^-(s)\bigr)\\
 = F_1(st^2) + F_2(s) + G_1(st) + G_2(t)
\end{multline}
with suitable functions $F_1$, $F_2$, $G_1$ and $G_2$ 
(cf.~\cite[Proposition~2.4]{Oc}).
Thus we have the following proposition.
\begin{prop}
For the functions $(U^{\pm}, V, W, F_1,F_2,G_1,G_2)$ satisfying \eqref{eq:b2fun}
we have the commuting differential operators \eqref{eq:b2-0} and \eqref{eq:b2-1} 
by putting
\begin{equation}
 \begin{cases}
   u^{\pm}(t) =\p_tU^{\pm}(e^t)+C',\ v(t)=\p_tV(e^t)+C,\ w(t)=\p_tW(e^t)+C,\\
   T(x,y) = \dfrac12\Bigl(\p_x^2-\p_y^2\Bigr)
            \Bigl(V(e^x)\bigl(U^+(e^{x+y})+U^-(e^{x-y})\bigr) - G_1(e^x)\Bigr)\\
            \quad\quad\quad\quad
            +C\bigl(u^+(x+y)+u^-(x-y)\bigr),\\
  \quad\quad C,\ C'\in\mathbb C.
 \end{cases}
\end{equation}
\end{prop}

Now we put
\begin{equation}
\begin{aligned}
\mathcal S(B_2)&=\bigl\{\bigl(U^+(t),U^-(t),V(t), W(t)\bigr);\,U^\pm, V \text{ and }W 
  \text{ are meromorphic}\\
 &\text{in a neighborhood of  0 and they satisfy }\eqref{eq:b2fun}\bigr\}.
\end{aligned}
\end{equation}

\begin{rem}\label{rem:b2-0}
 i)
Since the constant terms $U^{\pm}_0$, $V_0$ and $W_0$ have no effect on 
the equation \eqref{eq:b2i} and on the original functions 
$u^\pm$, $v$ and $w$, we will identify two functions appeared in the solutions of 
\eqref{eq:b2i} if they only differ in their constant terms.

ii) If $\bigl(U^+(t),U^-(t)\bigr)=0$ or $\bigl(W(t),V(t)\bigr)=0$, then \eqref{eq:b2i} is always true.
We call such $(U^+,U^-,W,V)\in \mathcal S(B_2)$ a {\it trivial solution\/} of 
\eqref{eq:b2fun}.
\end{rem}

We summarize elementary transformations acting on $\mathcal S(B_2)$.

\begin{lem}\label{rem:b2}
Let $\bigl(U^+(t),U^-(t),V(t),W(t)\bigr) \in\mathcal S(B_2)$.

{\rm i) (dual)}
$\bigl(V(t), W(t), U^+(t^2), U^-(t^2)\bigr)\in\mathcal S(B_2)$.

{\rm ii) (bilinear)}
If $\bigl(U^+(t),U^-(t),S(t), T(t)\bigr)\in\mathcal S(B_2)$, then
$\bigl(aU^+(t), aU^-(t), bV(t)+cS(t), bW(t)+cT(t)\bigr)\in\mathcal S(B_2)$ for 
$a,b,c\in\mathbb C$.

{\rm iii) (translations)}
$\bigl(U^+(ab^2t), U^-(bt), V(abt), W(at)\bigr)\in\mathcal S(B_2)$ for 
$a,b\in\mathbb C\setminus\{0\}$.

{\rm iv) (scaling)}
If $\bigl(U^+(t^r),U^-(t^r),W(t^r),V(t^r)\bigr)$ is well-defined for a suitable $r\in\mathbb Q\setminus\{0\}$, it is in $\mathcal S(B_2)$.

{\rm v) (symmetry)}
If $W(t)$ is a rational function, the reflection $(x,y)\mapsto(x,-y)$ can 
be applied to the solution and then 
$\bigl(U^-(t),U^+(t),V(t),-W(t^{-1})\bigr)\in\mathcal S(B_2)$.

{\rm vi) (symmetry)}
If $U^-(t)$ is a rational function, the reflection $(x,y)\mapsto(y,x)$ can 
be applied to the solution and then 
$\bigl(U^+(t),-U^-(t^{-1}),V(t),W(t)\bigr)\in\mathcal S(B_2)$.
\end{lem}
The lemma is a direct consequence of the definition of $\mathcal S(B_2)$.
For example, i) follows from
\begin{multline*}
 U^+(t^2s^2)\bigl(V(ts^2)+W(t)\bigr)+U^-(s^2)\bigl(V(ts^2)-W(t)\bigr)\\
 =  F_2(t) + F_1(s^2) + G_2(t^2s^2) + G_1(ts^2).
\end{multline*}

Note that the transformation in Lemma~\ref{rem:b2}~vi) is equals to a certain
composition of transformations in Lemma~\ref{rem:b2}~i), iv) and v).

\begin{defn}
If a solution of \eqref{eq:b2fun} obtained by applying transformations in 
Lemma~\ref{rem:b2} to an original solution, it is called a 
{\it standard transformation\/} of the original solution.
\end{defn}

We will study non-trivial solutions of \eqref{eq:b2fun}.
Considering standard transformations, we may assume
\begin{equation}
  (U^+_r,U^-_r) = (1,1) \text{ or } (1,0) \text{ or } (0,1).
\end{equation}

\begin{prop}\label{prop:rat}
Suppose $\bigl(U^+(t),U^-(t),V(t),W(t)\bigr)$ is a non-trivial solution of \eqref{eq:b2fun} with \eqref{b2:series2}.

{\rm i)} $U^\pm(t)$, $V(t)$ and $W(t)$ are rational functions.

{\rm ii)} {\rm (}\cite[Theorem~2.3]{Oc}{\rm )}
If $W(t)$ has a pole at $t=1$, then $U^+(t)=U^-(t)$
and $W(t^{-1})+W(t)=0$.
If $U^-(t)$ has a pole at $t=1$, then $V(t)=W(t)$ and $U^-(t^{-1})+U(t)=0$.

Here we note that this equality is interpreted in the sense of 
Remark~\ref{rem:b2-0}.

{\rm iii)} {\rm (}\cite[Corollary~3.8]{Oc}{\rm )} If at least two of
$\{U^+(t),U^-(t),V(t),W(t)\}$ have poles in $\mathbb C\setminus\{0\}$,
$(U^+,U^-,V,W)$ is a standard transformation of a solution given in the list
\eqref{eq:list-top} -- {\rm (Toda-$D_2^{(1)}$-S-bry)}.
\end{prop}
\begin{proof} i)
The equation \eqref{eq:b2i} shows $W_{q-2r}U^+_r=W_qU^-_r$ if $q>2|r|+|r'|$.
Hence $W(t)$ is a rational function and therefore so are $U^-(t)$, $U^+(t)$
and $V(t)$ because of Lemma~\ref{rem:b2}~i) and v).
\end{proof}

\begin{lem}\label{lem:b2-1}
{\rm i)} If $V(t)$ has a pole at the origin, then $U^+(t)$ and $U^-(t)$ are holomorphic
at the origin.

{\rm ii)} If $U^+(t)$ has a pole at the origin, then $V(t)$ and $W(t)$ are holomorphic
at the origin.
\end{lem}
\begin{proof}
If $r<0$ and $r'<0$ with $V_{r'}\ne0$, the coefficients of $s^{r+r'}t^{r'}$ and that of $s^{r+r'}t^{2r+r'}$ in \eqref{eq:b2fun} show $V_{r'}U^-_r=V_{r'}U^+_r=0$, 
which contradicts to $(U^+_r,U^-_r)\ne 0$.
Thus we have i) and then ii) by Lemma~\ref{rem:b2}~i).
\end{proof}

\begin{thm}\label{thm:B2}
Any non-trivial solution of \eqref{eq:b2fun} corresponds to a standard transformation
of a solution in the list \eqref{eq:list-top} -- \eqref{eq:list-last}.
\end{thm}
\begin{proof}
We will prove this theorem divided into several cases.

{\bf Case 1}: One of $U^+, U^-,V, W$ is zero.\newline
Proposition~\ref{prop:rat} assures that we may suppose $V=0$.
Then \eqref{eq:b2i} turns into
\begin{equation}\label{eq:W0}
 pq(2p-q)(p-q)(W_{q-2p}U^+_p - W_qU^-_p)=0.
\end{equation}

{\bf Case 1-1}: $V=0$, $W_{r''}\ne0$ and $(U^+_r,U^-_r)=(1,1)$.\newline
Suppose $\bar r:=-r>0$.
Then \eqref{eq:W0} with $p=-\bar r$ and $q=r''-2\bar r$
shows \[\bar r(r''-2\bar r)r''(r''-\bar r)W_{r''}U^+_r=0\]
and hence $r''=\bar r$ or $2\bar r$.
Since $\bar rq(2\bar r+q)(\bar r+q)(W_{q+2\bar r}U^+_r-W_qU^-_r)=0$,
\begin{equation}\label{eq:V0W}
 W(t)=at^{\bar r}(1-t^{\bar r})^{-1} + bt^{2{\bar r}}(1-t^{2\bar r})^{-1}
\end{equation}
Since $W(t)=at^{\bar r}(1+t^{\bar r})^{-1}$ if $2a+b=0$, we may assume
$W(t)$ has a pole at $t=1$ by applying a transformation in Lemma~\ref{rem:b2}~iii)
and hence $U^+(t)=U^-(t)$ by  Proposition~\ref{prop:rat}~ii).

On the other hand, \eqref{eq:W0} with $q=r''$ and that with $q=2p+r''$ show
\[
\begin{cases}
 pr''(2p-r'')(p-r'')W_{r''}U^-_p=0&\text{ for }p>0,\\
 p(2p+r'')r''(p+r'')W_{r''}U^+_p=0&\text{ for }p<0.
\end{cases}
\]
Thus we can conclude
\begin{equation}\label{eq:W0U}
 U^+(t)=U^-(t)=ct^{-\bar r}+dt^{-\frac{\bar r}2}+
 et^{\bar r}+ft^{\frac{\bar r}2}\quad\text{ with }bd=bf=0
\end{equation}
because $\bigl(U^+(t),U^-(t),0,bt^{2{\bar r}}(1-t^{2\bar r})^{-1}\bigr)\in\mathcal S(B_2)$.

If $r>0$, $rr''(2r-r'')(r-r'')W_{r''}U^-_r=0$ and
therefore $r''=2r$ or $r''=r$. Then by putting $\bar r=r$, 
the equation \eqref{eq:W0} with $p=r$ and $q> r''$ implies
\eqref{eq:V0W} and hence the same argument as above 
proves \eqref{eq:W0U}.

Hence the solution corresponds to a standard transformation of Case IV.

{\bf Case 1-2}: $V=0$ and $(U^+_r,U^-_r)=(1,0)$ or $(0,1)$.\newline
We have $q(2r-q)(r-q)W_q=0$ or $pq(2r-q)(r-q)W_{q-2r}=0$.
Hence $W(t)=at^{\bar r}+bt^{2\bar r}$ with $b\ne0$ and 
$\bar r\in\mathbb Z\setminus\{0\}$

If $a=0$, then \eqref{eq:W0} with $q=2\bar r$ implies
$U^-_p=0$ for $p\ne0$, $\bar r$, $2\bar r$.
If $a\ne0$, then \eqref{eq:W0} with $(p,q)=(\frac{\bar r}2,3\bar r)$ implies
$U^+_{\frac{\bar r}2}=0$, that with $q=2\bar r$
implies $U^-_p=0$ for $p\ne0$, $\bar r$, $2\bar r$
and that with $q=\bar r$ implies $U^-_p=0$ for $p\ne0$, 
$\pm\frac{\bar r}2$, $\bar r$.
Hence $U^-(t)=c^-t^{\bar r}+d^-t^{2\bar r}$ with $ad^-=0$.

Since $\bigl(U^+(t^{-1}),U^-(t^{-1}),0,W(t^{-1})\bigr)\in\mathcal S(B_2)$,
we have $U^+(t)=c^+t^{-\bar r}+d^+t^{-2\bar r}$ with $ad^+=0$
and the solution corresponds to the standard transform of Case V.

Now we may assume that none of $U^\pm(t),V(t),W(t)$ is zero and
\begin{equation}
 V_{r'}\ne0\quad\text{and}\quad W_{r''}\ne0.
\end{equation}
Lemma~\ref{lem:b2-1} and Lemma~\ref{rem:b2}~i) assure that we may assume 
$r>0$ except for the following case.

{\bf Case 2}: $W$ and $U^-$ have poles and $V$ and $U^+$ are holomorphic at the origin.\newline
Then $r''<0$ and $r<0$.
Put $\bar r=-r$.
The coefficients of $s^{-\bar r}t^q$ in \eqref{eq:b2fun} 
imply $W(t) = at^{-2\bar r} + bt^{-\bar r}$.

{\bf Case 2-1}: $W(t)=t^{-2\bar r} + bt^{-\bar r}$.\newline
The coefficients of $s^pt^{-2\bar r}$ imply $U^-(s)=s^{-\bar r}$.
Note that $U^-(t^{-1})$ and $W(t^{-1})$ are holomorphic at the origin and
$\bigl(U^+(t^{-1}), U^-(t^{-1}), V(t^{-1}), W(t^{-1})\bigr)\in\mathcal S(B_2)$.
If $U^+(t^{-1})$ has a pole at the origin, Lemma~\ref{lem:b2-1} assures that 
$V(t^{-1})$ is holomorphic there.
Hence we may assume $r>0$ by using a transformation in Lemma~\ref{rem:b2}~i) 
if necessary.

{\bf Case 2-2}: $W(t) = t^{-\bar r}$.\newline
The coefficients $s^pt^{-\bar r}$ in \eqref{eq:b2fun} imply
$U^-(s)=s^{-\bar r}+cs^{-\frac{\bar r}2}$
and this case is reduced to the previous case by Lemma~\ref{rem:b2}~i).

Now we may assume 
\begin{equation}
  r>0.
\end{equation}

{\bf Case 3}: $r>0$ and $r'>0$.\newline
Putting $p=r$ in \eqref{eq:b2i}, we have
\begin{equation}\label{eq:b2top}
  q(2r-q)(r-q)(W_{q-2r}U^+_r - W_qU^-_r)= 0.
\end{equation}

{\bf Case 3-1}: $(U^+_r,U^-_r)=(1,0)$ or $(0,1)$.\newline
Owing to Lemma~\ref{rem:b2}~v), we may assume $U^-_r=0$.
Then \eqref{eq:b2top} with $q=r''+2r$ means $r''=-2r$ or $r''=-r$
and \eqref{eq:b2i} with $q=r''$ means $U^-=0$.
Hence this case is reduced to Case 1.

{\bf Case 3-2}: $(U^+_r,U^-_r)=(1,1)$.\newline
The equation \eqref{eq:b2top} with $q=r''$ means $r''=r$ or $r''=2r$.

Note that \eqref{eq:b2top} means $W(t)=at^r(1-t^r)^{-1} + bt^{2r}(1-t^{2r})^{-1}$.
Since $U^\pm$, $V$ and $W$ are holomorphic at the origin, 
Lemma~\ref{rem:b2}~i) assures that if $r'\ne r''$,
this case is reduced to Case 3-1.
Hence we may assume $r'=r''$ and therefore $(V_{r'},W_{r'})=(1,1)$
by a suitable translation $s\mapsto as$.
It also follows from Lemma~\ref{rem:b2}~i) that
$U^-(s)=cs^{\frac {r'}2}(1-s^{\frac {r'}2})^{-1} + ds^{r'}(1-s^{r'})^{-1}$
and then this case is reduced to Proposition~\ref{prop:rat}~iii).

{\bf Case 4}: $r>0$ and $r'<0$.\newline
Using the transformation in Lemma~\ref{rem:b2}~v) if necessary, we
may assume 
\begin{equation}\label{eq:U-delta}
  (U^+_r,U^-_r)=(\delta,1)
\end{equation}
with $\delta=0$ or 1.
The equation \eqref{eq:b2i} with $p=r+m$, $q-p=\pm r$ and $m<0$ means
$(r+m)(2r+m)U^{\pm}_rV_m = 0$
and therefore
\begin{equation}
\begin{aligned}
   V(t) &= at^{-2r} + bt^{-r} + \sum_{j>0}V_jt^j,\\
   r' &= 
    \begin{cases}
       -2r\quad&\text{if }a\ne 0,\\
       -r\quad&\text{if }a = 0.\\
    \end{cases}
\end{aligned}
\end{equation}
Here $(a,b)\ne(0,0)$.
Moreover \eqref{eq:b2i} with $p=r$ shows
\begin{equation}\label{eq:b2negW}
 U^+_rW_{q-2r} = U^-_rW_q\quad
\text{ if }q\notin\{-2r,-r,0,r,2r,3r,4r\}.
\end{equation}
Put $\bar r=-\max\{r',r''\}>0$.
Then \eqref{eq:b2i} with $q=-\bar r$ means
\begin{equation}\label{eq:b2U}
  p\bar r(2p+\bar r)(p+\bar r)(V_{-\bar r}U^-_{p+\bar r} - W_{-\bar r}U^-_p)=0.
\end{equation}
Similarly \eqref{eq:b2i} with $q=-r$ means
\begin{equation}\label{eq:b2-41}
  p\bar r(2p+r)(p+r)(bU^-_{p+r} - W_{-r}U^-_p)=0.
\end{equation}

If $r' > r''$, we have $U^-=0$ because $V_{-\bar r}=0$.
Hence $r'\le r''$ and we may moreover assume
\begin{equation}\label{eq:VW-delta}
  (V_{-\bar r}, W_{-\bar r})=(1,\epsilon) \quad\text{with }
    \epsilon=0\text{ or }1\quad\text{and}\quad
    \bar r=\begin{cases}
             2r&\text{if }a\ne0,\\
             r&\text{if }a=0.
           \end{cases}
\end{equation}
Then \eqref{eq:b2negW} implies
\[
 W(t) = e_1t^r(1-\delta t^r)^{-1} + e_2t^{2r}(1-\delta t^{2r})^{-1}
     + e_3t^{-2r}+ e_4t^{-r} + e_5t^{r} + e_6t^{2r} + e_7t^{3r} + e_8t^{4r}
\]
and it follows from \eqref{eq:U-delta}, \eqref{eq:b2U} and \eqref{eq:VW-delta} that
\begin{equation}\label{eq:U-minus}
 U^-(s) = s^r(1-\epsilon s^r)^{-1} + cs^{2r}(1-\epsilon s^{2r})^{-1}.
\end{equation}
If $\epsilon=1$, we may assume that $U^-(s)$ has a pole at $s=1$ as 
in the argument in Case 1-1 and then $b=W_r$ in \eqref{eq:b2-41}.  
Note that if $b\ne0$, \eqref{eq:b2-41} implies $c=0$.
Hence
\begin{equation}\label{eq:bc0}
 bc=0
\end{equation}
and \eqref{eq:b2-41} with $p=r$ means
\begin{equation}
 e_4=0\quad\text{if }\epsilon=0.
\end{equation}

{\bf Case 4-1}: $(e_1,e_2)\ne0$ and $\delta=1$.\newline
We may also assume $W(t)$ has a pole at $t=1$.
If $\epsilon=1$, then $U^-(t)$ and $W(t)$ have poles in $\mathbb C\setminus\{0\}$
and this case is reduced to Proposition~\ref{prop:rat}~iii).
Hence we may assume $\epsilon=0$ and therefore $e_3=0$ by \eqref{eq:VW-delta}.
Then Proposition~\ref{prop:rat}~ii) assures 
\begin{equation}
\begin{aligned}
 W(t)&=e_1t^r(1- t^r)^{-1} + e_2t^{2r}(1- t^{2r})^{-1},\\
  U^+(s)&=U^-(s)=s^r+cs^{2r}.
\end{aligned}
\end{equation}
Now \eqref{eq:b2i} with $(p,q)=(2r,r)$ means $ce_1=0$.
Thus $(U^+(t),U^-(t),V(t),0)\in\mathcal S(B_2)$ and
it follows from Case 1 that $V(t)=at^{-2r} + bt^{-r}$ with $bc=0$.
Then the solution corresponds to (Toda-$B_2^{(1)}$-bry) or (Toda-$B_2^{(1)}$-S-bry).

{\bf Case 4-2}. $e_1=e_2=0$, $\epsilon=1$.\newline
Proposition~\ref{prop:rat}~ii) assures 
$V(t)=W(t)=at^{-2r}+ bt^{-r} + e_5t^{r} + e_6t^{2r} + e_7t^{3r} + e_8t^{4r}$
with $bc=0$.
Putting $q=2p-{\bar r}$ in \eqref{eq:b2i} we have 
$V_{\bar r}U^+_{p-\bar r}+W_{-\bar r}U^+_p=0$ if $p$ is sufficiently large positive 
integer.  
Hence if $U^+(t)$ is not a polynomial of $t$, 
it has a pole in $\mathbb C\setminus\{0\}$ 
and this case is reduced to Proposition~\ref{prop:rat}~iii).

Thus we may assume $U^+(s)=\sum_{i=r}^NU^+_is^i$ with $U^+_N\ne0$.
Suppose $W_j=0$ for $j>M$.
If $M>0$, the coefficients of $s^Nt^{M+2N}$ in \eqref{eq:b2fun} implies 
$W_MU^+_N=0$.  Hence $e_5=e_6=e_7=e_8=0$ and therefore $(U^+,0,V,W)\in\mathcal S(B_2)$
and Case 1 implies $U^+(s)=s^r+c's^{2r}$ with $bc'=0$.
The solution is a standard transform of case V.

{\bf Case 4-3}: $e_1=e_2=\epsilon=0$.\newline
Then $U^-(s)=s^r + cs^{2r}$ and 
$W(t)=e_5t^{r} + e_6t^{2r} + e_7t^{3r} + e_8t^{4r}$.
Putting $p=q+r$ in \eqref{eq:b2i}, we have
$V_qU^-_r-W_qU^-_{q+r}=0$ for $q>0$ and therefore
$V(t)=at^{-2r}  + b t^{-r} + ce_5t^r$.

Suppose $U^+(s)$ is not a polynomial of $s$.
Putting $q=2p+\bar r$ in \eqref{eq:b2i}, we have 
$U^+_{p+\bar r}+W_{\bar r}U^+_p=0$
for a sufficiently large $p$.  
We may assume $U^+(s)$ has a pole at $s=1$.
Then $W_{\bar r}=-1$ and Proposition~\ref{prop:rat}~ii)
with Lemma~\ref{rem:b2} proves $V(t^{-1})+W(t)=0$
and $V(t)=at^{-2r}+bt^{-r}$ with $bc=0$.
Thus $(U^+,0,V,W)\in\mathcal S(B_2)$ and 
$U^+(s)=d_1s^r(1-s^{r})^{-1} + d_2s^{2r}(1-s^{2r})^{-1}$ 
with $bd_2=0$
and the solution is a standard transform of Case IV.

Now we may assume $U^+(s)=\sum_{i=r}^NU^+_is^i$ and
$W(t)=\sum_{i=r}^MW_it^i$ with $W_MU^+_N\ne0$.
Then \eqref{eq:b2i} 
with $(p,q)=(N, M+2N)$ shows $W_MU^+_N=0$, which contradicts
to the assumption.
\end{proof}
\begin{cor}\label{cor:regA}
The non-trivial solutions $R(x,y)$ of \eqref{eq:b2-0} with regular 
singularity at the point $t=0$ are transformations of the following solutions
under translations. 
\begin{align*}
&\begin{aligned}
 &C_1\bigl(\sh^{-2}\lambda(x+y) + \sh^{-2}\lambda(x-y)\bigr)
+C_2\bigl(\sh^{-2}\lambda x + \sh^{-2}\lambda y\bigr)\\
&\quad+C_3\bigl(\sh^{-2}2\lambda x + \sh^{-2}2\lambda y\bigr)
+C_0,
\end{aligned}\tag{\rm Trig-$BC_2$-reg}
\allowdisplaybreaks\\
&\begin{aligned}
 &C_1\bigl(\sh^{-2}\lambda(x+y) + \sh^{-2}\lambda(x-y)\bigr)\\
 &\quad+C_2\bigl(\sh^{-2}2\lambda(x+y) + \sh^{-2}2\lambda(x-y)\bigr)\\
 &\quad+C_3\bigl(\sh^{-2}2\lambda x + \sh^{-2}2\lambda y\bigr)
 +C_0,
\end{aligned}\tag{\rm Trig${}^d$-$BC_2$-reg}
\allowdisplaybreaks\\
&C_1\bigl(e^{-2\lambda(x+y)}+e^{-2\lambda(x-y)})
+C_2\sh^{-2}\lambda y +C_3\sh^{-2}2\lambda y
+C_0,\tag{\rm Toda-$D_2$-bry}
\\
&C_1\sh^{-2}\lambda(x-y) +C_2\sh^{-2}2\lambda(x-y)
+ C_3\bigl(e^{-4\lambda x}+e^{-4\lambda y})
+C_0,\tag{Toda${}^{d}$-$D_2$-bry}
\allowdisplaybreaks\\
&\begin{aligned}
 C_1\sh^{-2}\lambda(x-y)
+C_2\bigl(e^{-2\lambda x}+e^{-2\lambda y})
+C_3\bigl(e^{-4\lambda x}+e^{-4\lambda y})
+C_0,
\end{aligned}\tag{\rm Trig-$A_1$-bry-reg}
\allowdisplaybreaks\\
&\begin{aligned}
 &C_1\bigl(e^{-\lambda(x+y)}+e^{-\lambda(x-y)})
+C_2\bigl(e^{-2\lambda(x+y)}+e^{-2\lambda(x-y)})\\
&\quad+C_3\sh^{-2}\lambda y
+C_0,
\end{aligned}\tag{\rm Trig${}^d$-$A_1$-bry-reg}
\allowdisplaybreaks\\
 &C_1e^{-\lambda(x-y)} + C_2e^{-\lambda y} + C_3e^{-2\lambda y}+C_0,
\tag{\rm Toda-$BC_2$}\\
 &C_1e^{-\lambda(x-y)} + C_2e^{-2\lambda(x-y)} + C_3e^{-2\lambda y}+C_0
\tag{\rm Toda${}^d$-$BC_2$}.
\end{align*}
\end{cor}

\section{Type $B_n$ ($n\ge3$)}\label{sec:Bn}
Let $\mathbb R^n$ be the Euclidean space with the natural inner product
$\langle x,y \rangle=\sum_{i=1}^nx_iy_i$
for $x=(x_1,\ldots,x_n)$, $y=(y_1,\ldots,y_n)\in\mathbb R^n$.
Then $e_i=(\delta_{i1},\ldots,\delta_{i\nu},\ldots,\delta_{in})$ for $i=1,\ldots,n$
form a natural orthonormal basis of $\mathbb R^n$.
For $v\in\mathbb R^n$, let $\p_v$ be the differential operator defined by 
$(\p_v\psi)(x)=\dfrac{d\psi(x+tv)}{dt}\Big|_{t=0}$ for a function $\psi(x)$ on 
$\mathbb R^n$ and we put $\p_i=\p_{e_i}$.
If $v\ne0$, the reflection $w_v$ with respect to $v$ is a linear transformation
of $\mathbb R^n$ defined by $w_v(x)=x -\dfrac{2\langle v,x\rangle}{\langle v,v\rangle}v$
for $x\in\mathbb R^n$.

The root system $\Sigma=\Sigma(B_n)$ of type $B_n$ is realized in $\mathbb R^n$ by
\begin{equation}
\begin{cases}
 \Sigma(A_{n-1})^+ = \{e_i-e_j;\, 1 \le i<j\le n\},\\
 \Sigma(D_n)^+ = \Sigma_L^+ = \{e_i \pm e_j;\, 1\le i<j\le n\},\\
 \Sigma(D_n) = \Sigma_L = \{\alpha,-\alpha;\,\alpha\in\Sigma(D_n)^+\},\\
 \Sigma(B_n)_S^+ = \Sigma_S^+ = \{e_k;\, 1\le k\le n\},\\
 \Sigma(B_n)^+ = \Sigma^+ = \Sigma(D_n)^+\cup\Sigma(B_n)^+_S,\\
 \Sigma(B_n) = \{\alpha,\,-\alpha;\,\alpha\in\Sigma(B_n)^+\}.
\end{cases}
\end{equation}
The Weyl group $W_\Sigma$ of $\Sigma$ is the finite group generated by 
$\{w_\alpha;\,\alpha\in\Sigma\}$,
which is the group generated by the permutation of the coordinate 
$(x_1,\ldots,x_n)$ of 
$\mathbb R^n$ and by the change of the signs of some coordinates $x_i$.
For a subset $F$ of $\Sigma$, let $W_F$ denote the subgroup of $W_\Sigma$ generated by 
$\{w_\alpha;\,\alpha\in F\}$.  Then we call the set 
$\bar F=\{w\alpha;\,w\in W_F\text{ and }\alpha\in F\}$ the root system generated by 
$F$ and $W_F$ the Weyl group of the root system $\bar F$.
Let
\begin{equation}
 P=\sum_{j=1}^n\frac{\p^2}{\p x_j^2} + R(x)
\end{equation}
be a differential operator with a function $R(x)$ such that it admits a differential operator
\begin{equation}
 Q=\sum_{j=1}^n \frac{\p ^4}{\p x_j^4} + S
 \qquad\text{with }\ord S<4
\end{equation}
satisfying $PQ=QP$.

Now we assume
\begin{equation}\label{eq:u-alpha}
\begin{aligned}
 R(x) &= \sum_{\alpha\in\Sigma(B_n)^+}u_\alpha(\langle\alpha,x\rangle)\\
      &= \sum_{1\le i<j\le n}\bigl(u^{ij}(x_i+x_j)+v^{ij}(x_i-x_j)\bigr)
         + \sum_{k=1}^n w^k(x_k),\\
 u^{ij}&= u_{e_i-e_j},\ v^{ij}=u_{e_i+e_j}\quad\text{and}\quad w^k=u_{e_k}\\
 &\quad\text{for }1\le i<j\le n\text{ and }1\le k\le n.
\end{aligned}
\end{equation}
For $\alpha\in\Sigma(B_n)^+$, 
we put $u_{-\alpha}(t)=u_\alpha(-t)$ for the convention.

Fix indices $i$ and $j$ with $1\le i<j\le n$ and put $u^{ji}(t) = u^{ij}(-t)$
and $I(i,j)=\{1,\ldots,n\}\setminus\{i,j\}$.
It follows from the proof of \cite[Theorem~6.1]{OOS} that
the condition for the existence of $Q$ is equivalent to
\begin{equation}\label{eq:bn}
S_{ij}=S_{ji}\qquad(1\le i<j\le n)
\end{equation}
with
\[
\begin{aligned}
S^{ij} &= \Bigl(\p_i^2 w^i(x_i) + \sum_{\nu\in I(i,j)}\p_i^2\bigl(u^{i\nu}(x_i+x_\nu)
 +v^{i\nu}(x_i-x_\nu)\bigr)\Bigr)\\
&\quad\cdot\Bigl(u^{ij}(x_i+x_j)-v^{ij}(x_i-x_j)\Bigr)\\
&+ 3\Bigl(\p_i w^i(x_i) + \sum_{\nu\in I(i,j)}
 \p_i\bigl(u^{i\nu}(x_i+x_\nu)+v^{i\nu}(x_i-x_\nu)\bigr)\Bigr)\\
&\quad\cdot\Bigl(\p_iu^{ij}(x_i+x_j)-\p_iv^{ij}(x_i-x_j)\Bigr)\\
&+ 2\Bigl(w^i(x_i) + 
\sum_{\nu\in I(i,j)}\bigl(u^{i\nu}(x_i+x_\nu)+v^{i\nu}(x_i-x_\nu)\bigr)\Bigr)\\
&\quad\cdot\Bigl(\p_i^2u^{ij}(x_i+x_j)-\p_i^2v^{ij}(x_i-x_j)\Bigr)\\
&+ \sum_{\nu\in I(i,j)}\Bigl(\p_i^2u^{i\nu}(x_i+x_\nu)-\p_i^2v^{i\nu}(x_i-x_\nu)\Bigr)
  \Bigl(u^{j\nu}(x_j+x_\nu) - v^{j\nu}(x_j-x_\nu)\Bigr).
\end{aligned}
\]

Then we have assumed that
\begin{equation}
  u_\alpha(\log t)=\sum u^\alpha_\nu t^\nu\quad\text{for }\alpha\in\Sigma^+
\end{equation}
with $u^\alpha_\nu\in\mathbb C$.
Here $u_\alpha(\log t)$ is analytic if $0<|t|\ll 1$ and $u^\alpha_\nu=0$
if $\nu$ is a sufficiently large negative integer.

Put
\begin{equation}
\begin{aligned}
 t_j &= e^{-x_j + x_{j+1}}\ (j=1,\ldots,n-1),\quad t_n = e^{-x_n},\\
 u^{ij}(x_i+x_j)&
   =\sum u^{ij}_\nu t_i^\nu\cdots t_{j-1}^\nu t_j^{2\nu}\cdots t_n^{2\nu},\\
 v^{ij}(x_i-x_j)&
   =\sum v^{ij}_\nu t_i^\nu\cdots t_{j-1}^\nu,\\
 w^i(x_k)&
   =\sum w^i_\nu t_k^\nu\cdots t_n^\nu,\\
  U_\alpha(t) &= \sum_{\nu\in\mathbb Z\setminus\{0\}} U^\alpha_\nu t^\nu\quad\text{with }
   u^\alpha_\nu = \nu U^\alpha_\nu\text{ and }\alpha\in\Sigma^+.
\end{aligned}
\end{equation}
Here $1\le i<j\le n$ and $u^{ij}_\nu$, $v^{ij}_\nu$ and $w^i_\nu\in\mathbb C$
and they are zero if $\nu$ is a sufficiently big negative integer.
Then the coefficients of $(t_i\cdots t_{j-1})^q(t_j\cdots t_n)^p$
in \eqref{eq:bn} show
\[
   pqw^i_{2p-q}u^{ij}_{q-p} - p(2p-q)w^j_qv^{ij}_{p-q}
 = q(q-p)w^i_{q-2p}u^{ij}_p - (2p-q)(p-q)w^ju^{ij}_p
\]
if $pq(p-q)(2p-q)\ne0$ and therefore by putting
\begin{equation}
  \begin{cases}
    U^\pm(t) = \sum_{\nu\ge r}U^\pm_it^\nu,\ 
    V(t) = \sum_{\nu\ge r}V_\nu t^\nu \text{ and }
    W(t) = \sum_{\nu\ge r''}W_\nu t^\nu,\\
    U^\pm_0=V_0=W_0=0,\\
    u^{ij}(t)=t(U^+)'(t)+u^{ij}_0,\ v^{ij}(t)=t(V^-)'(t)+v^{ij}_0,\\
    w^i(t)=tV'(t)+w^i_0
    \text{ and }w^j(t)=tW'(t)+w^j_0,
  \end{cases}
\end{equation}
we have \eqref{eq:b2i} and \eqref{eq:b2fun}.
Hence $(u^{ij}, v^{ij}, w^i, w^j)$ is a standard transformation of a solution
of type $B_2$ studied in \S\ref{sec:B2}.

Suppose $\{\alpha,\beta,\alpha+\beta\}\subset\Sigma^+_L$.
Then $(\alpha,\beta,\alpha+\beta)$ is one of the followings
\begin{gather}
  (e_{i_1-i_2}, e_{i_2-i_3}, e_{i_1-i_3}),\label{eq:d3-1}\\
  (e_{i_1-i_2}, e_{i_2+i_3}, e_{i_1+i_3}),\label{eq:d3-2}\\
  (e_{i_2-i_3}, e_{i_1+i_3}, e_{i_1+i_2}),\label{eq:d3-3}\\
  (e_{i_1-i_3}, e_{i_2+i_3}, e_{i_1+i_2}) \label{eq:d3-4}
\end{gather}
with $1\le i_1<i_2<i_3\le n$.

Put $s=e^{-\langle\alpha,x\rangle}$ and $t=e^{-\langle\beta,x\rangle}$.
Moreover put $(i,j,\nu)=(i_1,i_2,i_3)$, $(i_1,i_2,i_3)$, $(i_2,i_3,i_1)$ and $(i_1,i_3,i_2)$
according to \eqref{eq:d3-1}, \eqref{eq:d3-2}, \eqref{eq:d3-3} and \eqref{eq:d3-4},
respectively.
Then $(u_\alpha,u_\beta,u_{\alpha+\beta})= (v^{ij},v^{j\nu},v^{i\nu})$, 
$(v^{ij},u^{j\nu},u^{i\nu})$, $(v^{ij},u^{j\nu},u^{i\nu})$ and 
$(v^{ij},u^{j\nu},u^{i\nu})$, respectively, and
the coefficients of $s^pt^q$ in \eqref{eq:bn} show
\begin{multline}
  (- q^2-3(p-q)q-2(p-q)^2)u^{\alpha+\beta}_qu^\alpha_{p-q} 
  + p^2u^{\alpha+\beta}_pu^\beta_{q-p}\\
 = (-q^2+3pq-2p^2)u^\beta_qu^\alpha_p + (q-p)^2u^{\alpha+\beta}_pu^\beta_{q-p}.
\end{multline}
if $pq(p-q)\ne0$.
Hence \begin{equation}\label{eq:bni1}
pq(p-q)(2p-q)(U^{\alpha}_pU^{\beta}_q - U^{\alpha}_{p-q}U^{\alpha+\beta}_q 
  - U^{\beta}_{q-p}U^{\alpha+\beta}_p)=0.
\end{equation}
Now put $(i,j,\nu)=(i_2,i_3,i_1)$, $(i_1,i_3,i_1)$, $(i_1,i_3,i_2)$ and $(i_2,i_3,i_1)$
according to \eqref{eq:d3-1}, \eqref{eq:d3-2}, \eqref{eq:d3-3} and \eqref{eq:d3-4}, 
respectively.
Then $(u_\alpha,u_\beta,u_{\alpha+\beta})= (v^{i \nu},v^{ij},v^{j\nu})$, 
$(v^{i\nu},u^{ij},u^{j\nu})$, $(v^{j\nu},u^{ij},u^{i\nu})$ and 
$(v^{j\nu},u^{ij},u^{i\nu})$, respectively,
and the coefficients of $s^pt^q$ in \eqref{eq:bn} show
\begin{multline}
  \bigl(p^2+3(q-p)p+2(q-p)^2\bigr)u^{\alpha+\beta}_pu^\beta_{q-p} 
  - q^2u^{\alpha+\beta}_qu^\alpha_{p-q}\\
 = (p^2-3pq+2q^2)u^\alpha_pu^\beta_q - (p-q)^2u^{\alpha+\beta}_qu^\alpha_{p-q}
\end{multline}
if $pq(p-q)\ne0$ and we have
\begin{equation}\label{eq:bni2}
pq(p-q)(p-2q)(U^{\alpha}_pU^{\beta}_q - U^{\alpha}_{p-q}U^{\alpha+\beta}_q 
  - U^{\beta}_{q-p}U^{\alpha+\beta}_p)=0.
\end{equation}
Combining \eqref{eq:bni1} and \eqref{eq:bni2}, we get
\begin{equation}\label{eq:bni0}
pq(p-q)(U^{\alpha}_pU^{\beta}_q - U^{\alpha}_{p-q}U^{\alpha+\beta}_q 
  - U^{\beta}_{q-p}U^{\alpha+\beta}_p)=0.
\end{equation}
Namely,
\begin{equation}\label{eq:A2B}
 \bigl(U_\alpha(s)+U_\beta(t)-U_{\alpha+\beta}(st)\bigr)^2=
  F_\alpha(s)+F_\beta(t)+F_{\alpha+\beta}(st)
\end{equation}
with suitable functions $F_\alpha$, $F_\beta$ and $F_{\alpha+\beta}$.
Then if at least two in $\{U_\alpha,U_\beta,U_{\alpha+\beta}\}$ do not
vanish, Proposition~\ref{prop:A} shows that
$\bigl(U_\alpha(t),U_\beta(t),U_{\alpha+\beta}(t)\bigr)$ is a standard 
transformation of 
$\bigl(t^r(1-t^r)^{-1},t^r(1-t^r)^{-1},t^r(1-t^r)^{-1}\bigr)$ or 
$(C_1t^r,C_2t^r,C_3t^{-r})$.

The argument above shows the following lemma.

\begin{lem}\label{lem:two}
Let $\alpha$ and $\beta\in\Sigma$ such that
$\alpha\ne\pm\beta$, $\langle\alpha,\beta\rangle\ne0$, $|\alpha|\ge|\beta|$,
$U_\alpha\ne0$ and $U_\beta\ne0$.
Suppose
\[
  U_\alpha(t)=C_1t^r.
\]
Then we have the following two cases.

{\bf Case 1:} $|\alpha|=|\beta|$.
\begin{equation*}
 U_\beta(t) =
 \begin{cases}
   C_1't^r\quad&\text{if }\langle\alpha,\beta\rangle<0,\\
   C_1't^{-r}\quad&\text{if }\langle\alpha,\beta\rangle>0.
 \end{cases}
\end{equation*}

{\bf Case 2:} $|\alpha|^2=2|\beta|^2$.
\begin{equation*}
   U_\beta(t)=C'_1t^r(1-t^r)+C'_2t^{2r}(1-t^{2r})\text{ and }
   U_{w_\beta(\alpha)}(t)=U_\alpha(t)\text{ under a translation}
\end{equation*}
or
\begin{equation*}
 U_\beta(t)=
 \begin{cases}
   C_1't^r+C_2't^{2r}\quad&\text{if }\langle\alpha,\beta\rangle<0,\\
   C_1't^{-r}+C_2't^{-2r}\quad&\text{if }\langle\alpha,\beta\rangle>0.
 \end{cases}
\end{equation*}
\end{lem}

\begin{defn}\label{def:comp}
For the functions $u_\alpha$ in \eqref{eq:u-alpha}, put
\begin{equation}
 \Delta = \{\alpha\in\Sigma(B_n)^+;\, u'_\alpha\ne0\}.
\end{equation}
Let $\bar\Delta$ be the root system generated by $\Delta$
and let
\begin{equation}
 \bar\Delta = \bar\Delta_1\cup\cdots\cup\bar\Delta_N
\end{equation}
be the decomposition into irreducible root systems.
Put
\begin{equation}
  \Delta_k = \bar\Delta_k\cap\Delta
\end{equation}
and we call it an {\it irreducible component\/} of $\Delta$.

We say that $P$ with the potential function $R(x)$ is {\it irreducible\/} if 
$\bar\Delta$ is an irreducible root system of rank $n$ or of type $A_{n-1}$.
\end{defn}
\begin{rem} {\rm i)}
$\Delta=\Delta_1\cup\cdots\cup\Delta_N$.

{\rm ii)}
 Suppose $\alpha\in\Delta_k$.  Then $\beta\in\Delta$ is an element of $\Delta_k$
if and only if there exists a sequence
$\alpha_1=\alpha$, $\alpha_2,\ldots,\alpha_\ell=\beta\in\Delta$ such that
$\langle\alpha_i,\alpha_{i+1}\rangle\ne0$ for $i=1,\ldots,\ell-1$.
\end{rem}

\begin{lem}\label{lem:long}
Let $\Delta'$ be an irreducible component of $\Delta$
and let $\bar\Delta'$ and $\bar\Delta'_L$ be the root systems generated 
by $\Delta'$ and $\Delta'_L:=\Delta'\cap\Sigma_L$, respectively.
Suppose $\bar\Delta'$ is of type $B_m$ with $m>2$.
Then $\bar\Delta'_L$ is of type $A_{m-1}$ or type $D_m$.
\end{lem}
\begin{proof}
Since $\Delta'_L$ and $\{e_1,\ldots,e_m\}$ generate $\Delta'$,
there exist $\alpha_i\in\Delta'_L$ such that
$\langle\alpha_i,e_i\rangle\ne0$ and 
$\langle\alpha_i,e_{i+1}\rangle\ne0$ for $1\le i< m$.
Since $\alpha_1,\ldots,\alpha_{m-1}$ generate a root system of type $A_{m-1}$,
$\bar\Delta'_L$ is of type $A_{m-1}$ or type $D_m$.
\end{proof}
\begin{lem}\label{lem:EXT}
Let $S$ be a subset of a classical root system $\tilde\Sigma$ of type $A$ or $B$.
Suppose $S$ generates an irreducible root system $\bar S$ and
\[
 \langle\alpha,\beta\rangle\le0\quad\text{for }\alpha\in S\text{ and }\beta\in S
 \text{ with }\alpha\ne\beta.
\]
If the rank of $\bar S$ is larger than one, 
$S$ is the image of one of the following sets under the suitable 
transformation by an element of the Weyl group of $\tilde\Sigma$.
\begin{align}
 &\{e_1-e_2,\ldots,\,e_{m-1}-e_m\}\quad&(m\ge 3),\label{eq:Am}\\
 &\{e_1-e_2,\ldots,\,e_{m-1}-e_m,e_m\}\quad&(m\ge 2),\label{eq:Bm}\\
 &\{e_1-e_2,\ldots,\,e_{m-1}-e_m,\,e_{m-1}+e_m\}\quad&(m\ge4),\label{eq:Dm}\\
 &\{e_1-e_2,\ldots,\,e_{m-1}-e_m,\,e_m-e_1\}\quad&(m\ge 3),\label{eq:EAm}\\
 &\{e_1-e_2,\ldots,\,e_{m-1}-e_m,\,e_m,\,-e_1\}\quad&(m\ge 2),\label{eq:EBm}\\
 &\{e_1-e_2,\ldots,\,e_{m-1}-e_m,\,e_{m-1}+e_m,
\,-e_1-e_2\}\quad&(m\ge 4),\label{eq:EDm}\\
 &\{e_1-e_2,\ldots,\,e_{m-1}-e_m,\,e_{m-1}+e_m,\,-e_1\}\quad&(m\ge 3).\label{eq:BDm}
\end{align}
\end{lem}
\begin{proof}
Let $S'$ be a subset of $S$ such that
$S'$ is a transformation of one of the sets $\{e_1\}$, $\{e_1-e_2\}$, 
\eqref{eq:Am}, \eqref{eq:Bm} and \eqref{eq:Dm} by an element of
the Weyl group of $\tilde\Sigma$.

If the number of the elements of $S'$ is smaller than the rank of $\bar S$,
there exists $\alpha\in S$ and $\beta\in S'$ such that
$\langle\alpha,\beta\rangle\ne0$ and 
$\alpha\notin\sum_{\gamma\in S'}\mathbb R\gamma$.
Then it is easy to see that $S'\cup\{\alpha\}$ is a transformation
of \eqref{eq:Am}, \eqref{eq:Bm} or \eqref{eq:Dm} by a suitable
element of the Weyl group.

Thus we may assume that $S'$ equals \eqref{eq:Am} or \eqref{eq:Bm} or \eqref{eq:Dm}
and that the number of the elements of $S'$ equals the rank of $\bar S$.
Put $S_o=\{\beta\in\tilde\Sigma\cap\sum_{\gamma\in S'}\mathbb R\gamma;
\langle\beta,\gamma\rangle\le0\ \text{for any }\gamma\in S'\}$.
Then if $S'$ is \eqref{eq:Am} or \eqref{eq:Bm}, 
$S_o=\{e_m-e_1\}$ or $\{-e_1, -e_1-e_2\}$, respectively.
If $S'$ is \eqref{eq:Dm}, then 
$S_o=\{-e_1-e_2,-e_1\}$ or $\{-e_1-e_2\}$ according to
$\tilde\Sigma$ is of type $B$ or $A$, respectively.
Thus the lemma is clear because $S'\subset S\subset S'\cup S_o$.
\end{proof}

\begin{lem}
Let $\bar\Delta$ be a root system generated by a classical root system $\tilde\Sigma$
of type $A_{n-1}$ or $B_n$.
Let $\bar\Delta=\bar\Delta_1\cup\cdots\cup\bar\Delta_{N'}\cup\cdots\cup\bar\Delta_N$
be an irreducible decomposition so that the rank of $\bar\Delta_i$ is larger 
than one for $1\le i\le N'$ and the rank of $\bar\Delta_i$ equals one for $N'<i\le N$.
Then under the transformation by an element of the Weyl group of $\Sigma$,
there exists a sequence of integers $0=n_0<n_1<n_2<\cdots<n_{N'}$ such that
$\bar\Delta_i$ is generated by
\begin{equation*}
 \begin{cases}
    \{e_{n_{i-1}+1}-e_{n_{i-1}+2},\ldots,\,e_{{n_i}-1}-e_{n_i}\}
       \quad&\text{if $\bar\Delta_i$ is of type A},\\
    \{e_{n_{i-1}+1}-e_{n_{i-1}+2},\ldots,\,e_{{n_i}-1}-e_{n_i},\, e_{{n-i}-1}+e_{n_i}\}
        \quad&\text{if $\bar\Delta_i$ is of type D},\\
        \{e_{n_{i-1}+1}-e_{n_{i-1}+2},\ldots,\,e_{{n_i}-1}-e_{n_i},\, e_{n_i}\}
        \quad&\text{if $\bar\Delta_i$ is of type B}
 \end{cases}
\end{equation*}
for $1\le i\le N'$. Moreover if $N'<i\le N$.
$\bar\Delta_i$ equals $\{\pm e_\nu\}$, $\{\pm (e_\nu-e_{\nu+1})\}$ or
$\{\pm (e_\nu+e_{\nu+1})\}$ for a suitable $\nu$ with $\nu>n_{N'}$.
\end{lem}
\begin{proof}
Note that $\{\alpha\in\tilde\Sigma;\,\langle e_1-e_2,\alpha\rangle=0\}$
is generated by 
\[
 \begin{cases}
 \{e_3-e_4,\ldots,\,e_{n-1}-e_n\}\quad&\text{if $\tilde\Sigma$ is of type $A_n$},\\
 \{e_3-e_4,\ldots,\,e_{n-1}-e_n,\,e_n\}\text{ and }\{e_1+e_2\}&\text{if $\tilde\Sigma$ is of type $B_n$}.
 \end{cases}
\]
Hence  the lemma is clear by the induction on $N$ if $N'=0$.

Suppose $N'>0$.
By the preceding lemma, we may assume that the fundamental system of $\Delta_1$ is
\eqref{eq:Am}, \eqref{eq:Bm} or \eqref{eq:Dm}.
Then $\{\alpha\in\tilde\Sigma;\,\langle e_1-e_2,\alpha\rangle=0\}$ is generated by
\[
  \begin{cases}
 \{e_{m+1}-e_{m+2},\ldots,\,e_{n-1}-e_n\}\quad&\text{if $\tilde\Sigma$ is of type $A_n$},\\
 \{e_{m+1}-e_{m+2},\ldots,\,e_{n-1}-e_n,\,e_n\}&\text{if $\tilde\Sigma$ is of type $B_n$}
 \end{cases}
\]
and the lemma is clear by the induction on $N'$.
\end{proof}

\begin{rem}\label{rm:bn}
i) Fix $\alpha\in\Delta'$.
Let $v\in\mathbb R^n$ with $\langle\alpha,v\rangle=0$.
Then $\p_vu_\alpha(\langle\alpha,x\rangle)=0$.

ii) If the rank of $\bar\Delta'$ equals one, 
$u_\alpha(t)$ for $\alpha\in\Delta'$ is any function.

iii) If the rank of $\bar\Delta'$ is larger than one, $U_\alpha(t)$ with $\alpha\in\Delta'$ 
are global meromorphic functions 
and therefore we may study $\{U_\alpha(t);\,\alpha\in\Delta'\}$
under the image of a transformation by the Weyl group.

iv) By the irreducible decomposition in Definition~\ref{def:comp} 
our classification reduces to the case when $P$ is irreducible.
\end{rem}
\begin{thm}\label{prop:bn}
Let $\Delta'$ be an irreducible  component of $\Delta$.
Then the potential function
$R_{\Delta'}(x):=\sum_{\alpha\in\Delta'}u_{\alpha}(\langle\alpha,x\rangle)$ 
is a transformation of a function in the following list with $m\ge2$ under 
a map generated by the Weyl group, translations and scalings of the 
coordinates (cf.~Lemma~\ref{rem:b2}).
\begin{gather*}
\intertext{{\rm Type $A_1$:}
If the rank of $\bar\Delta'$ equals $1$, 
$R_{\Delta'}(x)$ is an arbitrary function of $\langle\alpha,x\rangle$
with $\alpha\in\Delta'$.
This solution is called trivial.}
\intertext{{\rm Type $B_2$:}
A standard transform of the function in the list {\rm (Trig-$B_2$) -- (Toda-$C_2^{(1)}$)} 
in $\S\ref{sec:B2}$ with replacing $(x,y)$ by $(x_1,x_2)$.}
 \begin{split}
  \intertext{{\rm(Trig-$B_m$):} Trigonometric potential of type $B_m$:}
  &\sum_{1\le i<j\le m}C_0\bigl(\sh^{-2}\lambda(x_i+x_j)
  +\sh^{-2}\lambda(x_i-x_j)\bigr)\\
  &\quad+\sum_{k=1}^m\bigl(
    C_1\sh^{-2}2\lambda x_k + C_2\sh^{-2}\lambda x_k + C_3\ch2\lambda x_k
  + C_4\ch4\lambda x_k\bigr).
  \end{split}\label{eq:BnP}\allowdisplaybreaks\\
 \begin{split}
\intertext{{\rm(Trig-$A_{m-1}$-bry):} Trigonometric potential of type $A_{m-1}$ with boundary:}
 &\sum_{1\le i<j\le m}C_0\sh^{-2}\lambda(x_i-x_j)\\
 &\quad+ \sum_{k=1}^m\bigl(C_1e^{-2\lambda x_k} + 
   C_2e^{-4\lambda x_k}+C_3e^{2\lambda x_k} + C_4e^{4\lambda x_k}\bigr),
 \end{split}\label{eq:Anb}\allowdisplaybreaks\\
 \begin{split}
\intertext{{\rm(Toda-$B_m^{(1)}$-bry):} Toda potential of type $B_m^{(1)}$ with boundary:}
 &\sum_{i=1}^{m-1}C_0e^{-2\lambda(x_i-x_{i+1})}
 + C_0e^{-2\lambda(x_{m-1}+x_m)}
 + C_1e^{2\lambda x_1} + C_2e^{4\lambda x_1}\label{eq:EBnb}\\
 &\quad+ C_3\sh^{-2}\lambda x_m +  C_4\sh^{-2}2\lambda x_m,
 \end{split}\allowdisplaybreaks\\
\intertext{{\rm(Toda-$C_m^{(1)}$):} Toda potential of type $C_m^{(1)}$:}
 \sum_{i=1}^{m-1}C_0e^{-2\lambda(x_i-x_{i+1})}
 + C_1e^{2\lambda x_1} + C_2e^{4\lambda x_1}
 + C_3e^{-2\lambda x_m}+ C_4e^{-4\lambda x_m},\label{eq:ECn}
\allowdisplaybreaks\\
 \begin{split}
\intertext{{\rm(Toda-$D_m^{(1)}$-bry):} Toda potential of type $D_m^{(1)}$ with boundary:}
 &\sum_{i=1}^{m-1}C_0\bigl(e^{-2\lambda(x_i-x_{i+1})}
 + e^{-2\lambda(x_{m-1}+x_m)}
 + e^{2\lambda(x_1+x_2)}\bigr)\label{eq:EDnb}\\
 &\quad+ C_1\sh^{-2}\lambda x_m + C_2\sh^{-2}{2\lambda x_m}
  + C_3\sh^{-2}\lambda x_1 + C_4\sh^{-2}{2\lambda x_1},
 \end{split}
\intertext{{\rm(Toda-$A_{m-1}^{(1)}$):} Toda potential of type $A_{m-1}^{(1)}$:}
 \sum_{i=1}^{m-1}C_0e^{-2\lambda(x_i-x_{i+1})}
 + C_0e^{2\lambda (x_1-x_m)}\label{eq:EAn}.
\end{gather*}
\end{thm}
\begin{defn} We define some potential functions as specializations of the above.
 
(Trig-$A_{m-1}$): Trigonometric potential of type $A_{m-1}$ is (Trig-$A_{m-1}$-bry) with 
$C_1=C_2=C_3=C_4=0$.

(Toda-$B_m^{(1)}$): Toda potential of type $B_m^{(1)}$ is (Toda-$B_m^{(1)}$-bry) with $C_3=C_4=0$.

(Toda-$D_m^{(1)}$): Toda potential of type $D_m^{(1)}$ is (Toda-$D_m^{(1)}$-bry) with 
$C_1=C_2=C_3=C_4=0$.

(Toda-$D_m$-bry): Toda potential of type $D_m$ with boundary is (Toda-$B_m^{(1)}$-bry) with 
$C_1=C_2=0$.

(Toda-$A_{m-1}$): Toda potential of type $A_{m-1}$ is (Toda-$C_m^{(1)}$) with 
$C_1=C_2=C_3=C_4=0$.

(Toda-$BC_m$): Toda potential of type $B_m$ is (Toda-$C_m^{(1)}$) with $C_1=C_2=0$.

(Toda-$D_m$): Toda potential of type $D_m$ is (Toda-$B_m^{(1)}$-bry) with 
$C_1=C_2=C_3=C_4=0$.
\end{defn}
{\em Proof of Proposition~\ref{prop:bn}}.
We may assume that $\bar\Delta'$ is not of type $B_2$. 
Then Lemma~\ref{lem:long} says that for any elements $\alpha$ and $\beta$ of 
$\Delta'_L$, there exists a sequence $\alpha=\alpha_1,\alpha_2,\ldots,\alpha_k=\beta$ 
such that $\langle\alpha_i,\alpha_{i+1}\rangle\ne0$ and $\alpha_i\in\Delta'_L$ for 
$i=1,\ldots,m-1$.
Note that the number of elements of $\Delta'_L$ is larger than one.
Fix $\alpha\in\Delta'$.
Then lemma~\ref{lem:two} assures that 
$U_\alpha(t)=Ct^r(1-at^r)^{-1}$ with $a\in\mathbb C$.

{\bf Case 1}: $U_{\alpha}(t)=Ct^r$.\newline
Lemma~\ref{lem:two} proves that $U_{\beta}(t)=C_\beta t^{2\epsilon(\beta)r}$ 
for $\beta\in\Delta'_L$.  Here $\epsilon(\beta)=1$ or $-1$.
Then the set $S_L=\{\epsilon(\beta)\beta;\,\beta\in\Delta'_L\}$ satisfies the 
assumption of Lemma~\ref{lem:EXT} and therefore we may assume that $S_L$ equals 
\eqref{eq:Am}, \eqref{eq:Dm}, \eqref{eq:EAm} or \eqref{eq:EDm}
under the transformation of an element of the Weyl group, which correspond to 
(Toda-$A_{m-1}$), (Toda-$D_m$), (Toda-$A_{m-1}^{(1)}$) and 
(Toda-$D_m^{(1)}$), respectively, if $U_{e_i}(t)=0$ for $i=1,\dots,m$.
Suppose $U_{e_i}(t)\ne0$ with a suitable $i$ satisfying $1\le i\le m$.
Then Lemma~\ref{lem:two} shows that one of the the following two cases occurs,
from which the statement of the proposition is clear.

{\bf Case 1-1}: $U_{e_i}(t)=C't^r(1-at^r)+C''t^{2r}(1-at^{2r})$ with $a\ne0$.\newline
We may assume $a=1$ by a translation.
Lemma~\ref{lem:two} shows that 
then $U_{-e_i-e_{i+1}}(t)=U_{e_i-e_{i+1}}(t)$ and if $i>1$
and $U_{e_{i-1}-e_i}(t)=U_{e_{i-1}+e_i}(t)$ if $i<m$.
Therefore 
\[
 \begin{cases}
 i=1\text{ and $S_L$ equals \eqref{eq:EDm}}
      &\bigl(\Rightarrow\text{(Toda-$D_m^{(1)}$-bry)}\bigr)\\
  \text{\ or}\\
 i=m\text{ and $S_L$ equals \eqref{eq:Dm} or \eqref{eq:EDm}}
      &{\bigl(\Rightarrow\text{(Toda-$B_m^{(1)}$-bry) or (Toda-$D_m^{(1)}$-bry)}\bigr)}.\\
 \end{cases}
\]
\quad
{\bf Case 1-2}: $U_{e_i}(t)=C't^{\epsilon_ir}+C''t^{2\epsilon_ir}$
with $\epsilon_i=\pm1$.

\noindent
Lemma~\ref{lem:two} says $\epsilon_i\langle e_i,\alpha\rangle\le 0$ 
for $\alpha\in S_L$, only the following cases occur.
\[
 \begin{cases}
   i=1\text{ and $S_L$ equals \eqref{eq:Am} or \eqref{eq:Dm}}
   &{\bigl(\Rightarrow\text{(Toda-$C_m^{(1)}$) or (Toda-$B_m^{(1)}$-bry)}\bigr)},\\
   i=m\text{ and $S_L$ equals \eqref{eq:Am}}
   &{\bigl(\Rightarrow\text{(Toda-$C_m^{(1)}$)}\bigr)}.
 \end{cases}
\]
\quad
{\bf Case 2}: $U_\alpha(t)=C_\alpha t^r(1-a_\alpha t^r)^{-1}$ with $a_\alpha\ne0$.\newline
The argument just before Lemma~\ref{lem:two} says 
that the condition $U_\beta(t)\ne0$ and 
$U_\gamma(t)\ne0$ with $|\alpha|=|\beta|=|\gamma|$ means $U_{w_\beta(\gamma)}(t)\ne0$.
Hence $\{\pm\beta;\,\beta\in\Delta'_L\}$ is a root system of type $A_{m-1}$ or $D_m$.
We may assume $a_\alpha=1$ for its simple root $\alpha$
and hence $C_\alpha$ and $a_\alpha$ does not depend on $\alpha\in\Delta'_L$
because of \eqref{eq:A2B} and Proposition~\ref{rem:a2}.

{\bf Case 2-1}: $\bar\Delta'_L$ is of type $A$.\newline
Considering $\{U_{e_i+e_{i+1}}, U_{e_i-e_{i+1}}, U_{e_i}, U_{e_{i+1}}\}$, 
Theorem~\ref{thm:B2} shows 
$U_{e_i}(t)=U_{e_{i+1}}(t)=C_1t^{r}+C_2t^{2r}+C_3t^{-r}+C_4t^{-2r}$
and this case is reduced to (Trig-$A_{m-1}$-bry).

{\bf Case 2-2}: $\bar\Delta'_L$ is of type $D$.\newline
By the same consideration as in the previous case we may assume
$U_{e_\nu}(t)=C_1t^r(1-t^r)^{-1}+C_2t^{2r}(1-t^{2r})^{-1} + C_3(t^r-t^{-r})+C_4(t^{2r}-t^{-2r})$ for $\nu=1,\ldots,m$.
Hence this case is reduced to (Trig-$B_m$).
\qed

\begin{cor}\label{cor:regB}
The non-trivial solutions in Proposition~\ref{prop:bn} which have regular 
singularity at the point $t=0$ are in Corollary~\ref{cor:regA} or 
in the following list.
\begin{gather*}
\allowdisplaybreaks
 \begin{aligned}
  &\sum_{1\le i<j\le m}C_0\bigl(\sh^{-2}\lambda(x_i+x_j)
  +\sh^{-2}\lambda(x_i-x_j)\bigr)\\
  &\quad+\sum_{k=1}^m\bigl(
    C_1\sh^{-2}2\lambda x_k + C_2\sh^{-2}\lambda x_k\bigr),
  \end{aligned}
\tag{\rm Trig-$BC_m$-reg}
\allowdisplaybreaks\\
 \sum_{1\le i<j\le m}C_0\sh^{-2}\lambda(x_i-x_j)
 + \sum_{k=1}^m\bigl(C_1e^{-2\lambda x_k} + 
   C_2e^{-4\lambda x_k}\bigr),
\tag{\rm Trig-$A_{m-1}$-bry-reg}
\allowdisplaybreaks\\
 C_0\sum_{i=1}^{m-1}e^{-2\lambda(x_i-x_{i+1})}
 + C_0e^{-2\lambda(x_{m-1}+x_m)}
 + C_3\sh^{-2}\lambda x_m +  C_4\sh^{-2}2\lambda x_m,
\tag{\rm Toda-$D_m$-bry}\\
 C_0\sum_{i=1}^{m-1}e^{-2\lambda(x_i-x_{i+1})}
 + C_3e^{-2\lambda x_m}+ C_4e^{-4\lambda x_m}.
\tag{\rm Toda-$BC_m$}
\end{gather*}
\end{cor}
\begin{rem}\label{rm:reg}
We have a natural compactification $X$ of the space $\mathbb C^n$ of $t$
so that for every $w\in W_{\Sigma(B_n)}$ 
\[
 s^w_j = e^{-(x'_j-x'_{j+1})}\quad(j=1,\ldots,n-1),\quad s^w_n = e^{-x'_n}
 \quad\text{with }
 x'=wx
\]
gives a local coordinate system of $X$ and $t_j=s^e_j$ $(j=1,\dots,n)$.
Then the non-trivial potential functions $R(x)$ we have obtained is 
meromorphic on $X$.

If $R(x)$ is holomorphic at $(s^w_1,\dots,s^w_n)=0$ for any $w\in W_{\Sigma(B_n)}$, $R(x)$ is said to have {\it regular singularity at every infinity\/}.
In this case, our classification says that $R(x)$ is decomposed to the 
functions {\rm (Trig-$BC_m$-reg)} and {\rm (Trig-$A_m$)} which exactly
corresponds to Heckman-Opdam's potential function of classical type.
This gives a characterization of Heckman-Opdam's hypergeometric equations.
\end{rem}


\begin{thebibliography}{OSh}
\bibitem[BB]{BB}
H.\ W.\ Braden and J.\ G.\ B.\ Byatt-Smith,
{\it On a functional differential equation of determinantal type}, 
Bull.\ London Math.\ Soc.\ {\bf 31}(1999), 463--470.

\bibitem[BP]{BP}
V.\ M.\ Buchstaber and A.\ M.\ Perelomov,
{\it On the functional equation related to the quantum three-body problem},
Contemporary Mathematical Physics, AMS Transl.\ Ser.\ {\bf 175}(1996), 15--34.

\bibitem[I]{I} V.\ I.\ Inozemtsev,
{\it Lax representation with spectral parameter on a torus for 
Integrable particle systems}, Lett.\ Math.\ Phys.\ {\bf 17}(1989), 11--17.

\bibitem[IM]{IM} V.\ I.\ Inozemtsev and D.\ V.\ Meshcheryakov,
{\it Extensions of the class of integrable dynamical systems
connected with semisimple Lie algebras}, Lett.\ Math.\ Phys.\
{\bf 9}(1985), 13--18.

\bibitem[HO]{HO} G.\ J.\ Heckman and E.\ M.\ Opdam
{\it Root system and hypergeometric functions.\/ {\rm I}}, Comp.\ Math.\
{\bf 64}(1987), 329--352.

\bibitem[Oc]{Oc} H.\ Ochiai,
{\it Commuting differential operators of rank two},  Indag.\ Math.\ (N.S.)
{\bf 7}(1996), 243--255

\bibitem[OO]{OO} H.\ Ochiai and T.\ Oshima, 
{\it Commuting differential operators with $B_2$ symmetry}, 
Funkcial.\ Ekvac.\ {\bf 46}(2003), 297--336.

\bibitem[OOS]{OOS} H.\ Ochiai, T.\ Oshima and H.\ Sekiguchi,
{\it Commuting families of symmetric differential operators}, Proc.\ Japan Acad.\ 
{\bf 70 A}(1994), 62--66.

\bibitem[OP1]{OP1} M.\ A.\ Olshanetsky and A.\ M.\ Perelomov,
{\it Classical integrable finite dimensional systems related to Lie algebras},
Phys.\ Rep.\ {\bf 71}(1981), 313--400.

\bibitem[OP2]{OP2} ------------,
{\it Quantum integrable systems related to Lie algebras}, Phys.\ Rep.\ 
{\bf 94}(1983), 313--404.

\bibitem[O1]{O1} T.\ Oshima, 
{\it A realization of Riemannian symmetric spaces},
J.\ Math.\ Soc.\ Japan {\bf 30}(1978), 117--132.

\bibitem[O2]{O2} ------------,
{\it Completely integrable systems with a symmetry in coordinates}, 
Asian Math.\ J.\ {\bf 2}(1998), 935--956.

\bibitem[OS]{OS} T.\ Oshima and H.\ Sekiguchi,
{\it Commuting families of differential operators invariant under the action of 
a Weyl group}, J.\ Math.\ Sci.\ Univ.\ Tokyo
{\bf 2}(1995), 1--75.

\bibitem[Ru]{Ru} S.\ N.\ M.\ Ruijsenaas,
{\it Systems of Calogero-Moser type},
Proceedings of the 1994 CRM Banff Summer School `Particles and Fields',
CRM Series in Mathematical Physics, pp.\ 251--352, Springer, 1999.

\bibitem[Ta]{Ta} K.\ Taniguchi,
{\it On the symmetry of commuting differential operators with singularities 
along hyperplanes},
Int.\ Math.\ Res.\ Not.\ {\bf 36}(2004), 1845--1867.

\bibitem[vD]{vD} J.\ F.\ van Diejen,
{\it Integrability of difference Calogero-Moser systems}, J.\ Math.\ Phys.\ 
{\bf 35}(1994), 2983--3004.

\bibitem[vD2]{vD2} ------------,
{\it Difference Calogero-Moser systems and finite Toda chains}, J.\ Math.\ Phys.\ 
{\bf 36}(1995), 1299--1323.

\bibitem[Wa]{Wa} S. Wakida,
{\it Quantum integrable systems associated with classical Weyl groups},
master thesis presented to University of Tokyo, 2004.
\end{thebibliography}
\end{document}